\documentclass[aps,10pt,prd,a4paper,twocolumn,notitlepage,nofootinbib,showpacs]{revtex4-1}
\usepackage{amsmath}
\usepackage{graphicx}
\usepackage{color}
\usepackage[normalem]{ulem} 

\allowdisplaybreaks

\newcommand{\be}{\begin{equation}}
\newcommand{\ee}{\end{equation}}

\newcommand{\pa}{\partial}

\newcommand{\bea}{\begin{eqnarray}}
\newcommand{\eea}{\end{eqnarray}}
\newcommand{\ben}{\begin{eqnarray*}}
\newcommand{\een}{\end{eqnarray*}}

\begin{document}
 
\title{General-relativistic rotation laws in rotating fluid bodies }

\author{Patryk Mach}
\author{Edward Malec}
\affiliation{Instytut Fizyki im.~Mariana Smoluchowskiego, Uniwersytet Jagiello\'nski, {\L}ojasiewicza 11, 30-348 Krak\'{o}w, Poland} 

\begin{abstract}
We   formulate new general-relativistic extensions of Newtonian rotation laws for self-gravitating stationary fluids. They  have been used   to re-derive,  in the first post-Newtonian approximation,   the  well known geometric dragging of frames. We derive two other    general-relativistic weak-field effects  within rotating tori: the  recently discovered  dynamic anti-dragging   and a new effect that  measures the deviation from the Keplerian motion and/or the contribution of the fluids selfgravity.     One can use the rotation laws to study the uniqueness and the  convergence  of the post-Newtonian approximations, and the existence of the post-Newtonian limits. 
\end{abstract}

\pacs{04.20.-q, 04.25.Nx, 04.40.Nr, 95.30.Sf}

\maketitle 

\section{Introduction}

Stationary Newtonian hydrodynamic configurations are characterized by a variety of rotation curves. The  angular momentum per unit mass  $j$ can be any function of $r$, where $r$ is the distance from the rotation axis. Other restrictions arise from stability considerations \cite{Tassoul}. In contrast to that, for a long time  the only  known  rotation law in general-relativistic hydrodynamics had been that with $j$ as a linear function of the angular velocity. Recently Galeazzi, Yoshida and Eriguchi \cite{GYE}  have found  a nonlinear angular velocity profile, that may approximate the Newtonian monomial rotation curves $\Omega_0=w/r^\lambda $  in the nonrelativistic limit. In this paper we   define general-relativistic rotation curves $j=j(\Omega )$  that in the nonrelativistic limit exactly coincide with   $\Omega_0=w/r^\lambda $ ($0 \le \lambda  \le 2$, $\lambda \ne 1$).  We are able to obtain the general-relativistic Keplerian rotation law that possesses the first post-Newtonian limit (1PN) and  exactly encompasses  the solution corresponding to the massless disk of dust in the Schwarzschild spacetime.

\section{Hydrodynamical equations}
We recapitulate, following    \cite{komatsu}, the equations of general-relativistic hydrodynamics.
Einstein equations, with the signature of the metric $(-,+,+,+)$, read
\begin{equation}
R_{\mu \nu} -g_{\mu \nu }\frac{R}{2} = 8\pi \frac{G}{c^4}T_{\mu \nu },
\label{ee1}
\end{equation}
where $T_{\mu \nu }$ is the stress-momentum tensor.  The \emph{stationary} metric reads

\begin{align}
\label{metric}
d s^2 &=  -  e^{\frac{2\nu }{c^2} }(d x^0)^2
+r^2   e^{\frac{2\beta }{c^2} } \left( d \phi  -  \frac{\omega }{c^3}\left( r, z \right) d x^0\right)^2
\nonumber\\&\qquad
+e^{\frac{2 \alpha }{c^2} }   \left( dr^2 +   dz^2\right)    .
\end{align}
Here $x^0 =ct$ is the rescaled  time coordinate, and $r$, $z$, $\phi$ are cylindrical  coordinates.  
 We assume axial symmetry and employ the stress-momentum tensor  
\be 
\label{emD}
T^{\alpha\beta} = \rho (c^2+h)u^\alpha u^\beta + p g^{\alpha\beta},
\ee
where $\rho$ is the baryonic rest-mass density, $h$ is the  specific enthalpy,
and $p$ is the  pressure.  
The 4-velocity  {$u^\alpha  $}  is normalized, $g_{\alpha\beta}u^\alpha u^\beta=-1$.
 The coordinate (angular)  velocity reads ${\vec v}= \Omega \partial_\phi $, where $\Omega = u^\phi /u^t$.

We assume a barotropic equation of state $p = p(\rho)$. To be more concrete, one can take the  polytropic equation of state  $p(\rho ,S) = K(S) \rho^\gamma$,
where $S$ is the specific entropy of fluid. Then one has $h(\rho ,S) = K(S) \frac{\gamma}{\gamma-1}\rho^{\gamma-1}$. The entropy is assumed to be constant.

Define the square of the linear velocity
\be
V^2=r^2 \left( \Omega -\frac{\omega }{c^2}\right)^2 e^{2\left( \beta - \nu \right)/c^2}.
\ee
The potentials $\alpha$, $\beta$, $\nu$, and $\omega$ satisfy  equations that have been found by Komatsu, Eriguchi and Hachisu \cite{komatsu}.  They constitute an overdetermined, but consistent,  set of equations. The general-relativistic Euler equations are solvable, assuming  an  integrability condition --- that    the angular momentum per unit mass,
\be
j  =  u_\phi u^t= \frac{V^2}{\left( \Omega -\frac{\omega }{c^2}\right) \left( 1-\frac{V^2}{c^2}\right) },
\ee
depends only on the angular velocity $\Omega $; $ j\equiv j(\Omega )$. In such a case  the Euler equations reduce to  a general-relativistic integro-algebraic Bernoulli equation, that    embodies  the hydrodynamic information carried by the continuity equations  $\nabla_\mu T^{\mu \nu }=0$ and the baryonic mass conservation
$\nabla_\mu \left( \rho u^\mu \right) =0$.  It is given by the expression  
\be
\label{grBernoulli}
\ln \left( 1+\frac{h}{c^2}\right) +\frac{\nu }{c^2} +\frac{1}{2}\ln \left( 1-\frac{V^2}{c^2}\right) +\frac{1}{c^2}\int d\Omega j(\Omega ) =C.
\ee

\section{Rotation laws}

The general-relativistic rotation law employed in the literature \cite{Bardeen_1970, Butterworth_Ipser, nishida_eriguchi,nishida1, komatsu} has the form
\be
j(\Omega ) = A^2 ( \Omega_c -\Omega ),
\ee
where $A$  and $\Omega_c$ are  parameters. In the Newtonian limit and large $A$ one arrives at the rigid rotation, $\Omega = \Omega_c$, while for small $A$ one gets the constant angular momentum per unit mass.  A three-parameter expression for $j$ is proposed in \cite{GYE}.

Below we define  a new family of rotation laws,
\be
\label{momentum}
j(\Omega ) \equiv \frac{w^{1-\delta }  \Omega^\delta }{1-   \frac{ \kappa  }{   c^2 }  w^{1- \delta }\Omega^{1+\delta } +\frac{\Psi }{   c^2}  },
  \ee
 where $w$, $\delta , \kappa  $ and $\Psi $ are   parameters. 
The rotation curves $\Omega \left( r, z \right) $ ought to  be recovered from  the equation 
\be
\label{rotation_law}
  \frac{w^{1-\delta }  \Omega^\delta }{1-   \frac{ \kappa  }{   c^2 }  w^{1- \delta }\Omega^{1+\delta } +\frac{\Psi }{   c^2}  }  = \frac{V^2}{\left( \Omega -\frac{\omega }{c^2}\right) \left( 1-\frac{V^2}{c^2}\right) },  
\ee
For $\delta \neq -1$, the general-relativistic Bernoulli equation (\ref{grBernoulli}) acquires  a simple algebraic  form

\begin{eqnarray}
\nonumber
 \left( 1 + \frac{h}{c^2}\right)  e^{\nu /c^2}   \sqrt{1-\frac{V^2}{c^2}} \times & & \\
\left( 1 - \frac{ \kappa  }{   c^2 }  w^{1- \delta }\Omega^{1+\delta } +\frac{\Psi }{   c^2} \right)^{\frac{-1}{\left( 1+\delta \right) \kappa }} & = & C.
\label{algebraic_Bernoulli}
\end{eqnarray}
We shall explain now the meaning and status of the four constants $w$, $\delta , \kappa  $ and $\Psi $.   Assume that there exists the Newtonian limit (the zeroth order of the post-Newtonian expansion ---  0PN) of the rotation law. This yields

\be
\label{zeroth_rotation_law}
\Omega_0 = \frac{w}{r^\frac{2}{1- \delta }}.
\ee
Thus $w$ and $\delta $ can be obtained from the Newtonian limit.  Moreover, the  constant $w $ is any real number, while $\delta $ is nonpositive  --- due to the stability requirement \cite{Tassoul} --- and satisfies the bounds $-\infty \le \delta \le  0$  and $\delta \neq -1$. These  two constants can be given apriori  within the given range of values.  Let us remark at this point that the rotation law (\ref{momentum}), and consequently the Newtonian rotation   (\ref{zeroth_rotation_law}), applies
primarily to single rotating toroids  and toroids rotating around black holes. In the case of rotating stars one would have to construct a special differentially rotating law, with the aim to avoid singularity at the rotation axis. 

The two limiting cases $\delta  =0$  and $ \delta =-\infty $ correspond to the constant angular momentum per unit mass  ($\Omega_0 =w/r^2$)  and the rigid rotation ($\Omega =w$), respectively. The Keplerian rotation is related  to the choice of $ \delta =-1/3$ and $w^2=GM$, where $M$ is a mass \cite{MMP}. The case with $\delta = -1$ should be considered separately, but we expect that the reasoning will be  similar.

The values of $\kappa$ and $\Psi$ are problematical.     One possibility to get them  is to  apply the  PN  expansion. The rotation law in the PN  expansion scheme should not be given apriori, but is expected to build up --- in the subsequent orders of $c^{-2}$ ---  from the   Newtonian rotation law.  The Newtonian rotation curves are specified  arbitrarily, but the next PN corrections should be  defined uniquely. This is, however,  a well known property of the post-Newtonian expansions, that they are non-unique. Damour, Jaranowski and Sch\"afer \cite{DJS} demand that a test body rotating circularly in a Schwarzschild space-time satisfies exactly the Keplerian rotation law with $\Omega^2 =GM/R^3$, where $R$ is the areal radius. Inspired by this we impose a   {\bf Fixing Condition} (F-Condition thereafter)  --- that a rotating infinitely thin disk made of  dust in a Schwarzschild space-time  satisfies exactly the Bernoulli equation and the Keplerian rotation law.

Consider  a rotating, infinitely thin and weightless   disk of dust in the Schwarzschild geometry.
This is a textbook knowledge that there exists a stationary solution --- each particle of dust can move along a circular  trajectory of a radius $R$ with the angular velocity
$\Omega = \sqrt{GM/R^3}$.  We shall present this solution in conformal coordinates, using our formalism. The conformal Schwarzschild metric reads
$ds^2=-\Phi^2/f^2 \left( dx^0\right)^2 +f^4\left( dr^2+dz^2+r^2d\phi^2\right)$, where $\Phi = 1-GM/(2c^2\sqrt{r^2+z^2})$ and $f=1+GM/(2c^2\sqrt{r^2+z^2})$. The angular velocity is equal to the Keplerian velocity $\Omega^2 =GM/(\sqrt{r^2+z^2}^3f^6)$  and $R=\sqrt{r^2+z^2}f^2$.  The total  energy per unit mass $\Psi $ vanishes for a test dust. Let the disk lie on the $z=0$ plane
and assume the rotation law with $\delta =-1/3$ and $\kappa =3$:
\be
\label{rotation_law_dust}
  \frac{w^{4/3 }  \Omega^{-1/3}}{1-   \frac{ 3 }{   c^2 }  w^{4/3 }\Omega^{2/3 }    }  = \frac{V^2}{ \Omega   \left( 1-\frac{V^2}{c^2}\right) }.  
\ee
Here $V^2=\Omega^2 r^2 f^6/\Phi^2$.  Notice that $h=0$;  the enthalpy per unit mass vanishes for dust.  This is a simple exercise to show that $w=\sqrt{GM}$ and $\Omega^2 =GM/(r^3f^6)$ solve both Eq.~(\ref{rotation_law_dust}) and the Bernoulli equation (\ref{algebraic_Bernoulli}); the constant in (\ref{algebraic_Bernoulli}) equals to unity.

\section{ 1PN corrections to angular velocity}
{\bf Taking into account the above, we  shall prove  that  if   $\kappa =(1 - 3\delta )/(1 + \delta ) + \mathcal{O}(c^{-2})$ and $\Psi =4c_0 + \mathcal{O}(c^{-2})$, where $c_0$ is the Newtonian hydrodynamic energy per unit mass,   then  the exact solution satisfies the first post-Newtonian (1PN) equations.  }    We shall use the formalism  of \cite{komatsu} and the rotation law (\ref{rotation_law}), and recover most of the  results obtained in the 1PN approach employed in \cite{JMMP}. Notice that if $\delta =-1/3$, then $\kappa =3$ --- one recovers the  coefficient in front of $w^{4/3 }\Omega^{2/3 } $ in (\ref{rotation_law_dust}) that is required by the F-Condition.

The 1PN approximation  corresponds to the choice of metric exponents
$\alpha =\beta =-\nu =-U$ with  $|U|\ll c^2$ \cite{BDS}. Define $\omega \equiv r^{-2} A_\phi$. The spatial part of the metric 
\begin{align}
\label{metric1}
d s^2 &=  -  \left(1+\frac{2U}{c^2} +\frac{2U^2}{c^4}\right) (d x^0)^2
-2  c^{-3} A_\phi d x^0 d \phi \ 
\nonumber\\&\qquad
+\left(1-\frac{2U}{c^2}  \right)  \left( d r^2 + d z^2 + r^2 d \phi^2\right)    .
\end{align} 
is conformally flat.   
  
We split different quantities ($\rho$, $p$, $h$, $U  $, and $v^i $) into their Newtonian (denoted by subscript `0' and 1PN (denoted by subscript `1') parts.
E.g., for $\rho$, $\Omega$, $\Psi$, and $U$ this splitting reads
\begin{subequations}
\label{density_rotation}
\begin{align}
\rho &= \rho_0 + c^{-2} \rho_1,
\\[1ex]
 \qquad
\Omega  &= \Omega_0  + c^{-2} v_1^\phi, 
\\[1ex] 
\qquad
\Psi  &= \Psi_0  +   { \mathcal{O}(c^{-2})},
\\[1ex] 
\qquad
U& = U_0 + c^{-2} U_1.
\end{align}
\end{subequations}
Notice that, up to the 1PN order,   
\begin{equation}
\label{enthalpy}
\frac{1}{\rho} \partial_i p = \partial_i h_0 + c^{-2} \partial_ih_1
 {+ \mathcal{O}(c^{-4})},
\end{equation}
where the {1PN} correction $h_1$ to the specific enthalpy can be written as $h_1 = \frac{dh_0}{d\rho_0} \rho_1$. For the polytropic equation of state this gives $h_1= \left( \gamma -1 \right) h_0 \rho_1 / \rho_0$.
 
Making use of the introduced above splitting of quantities into Newtonian 0PN and 1PN parts
one can extract from Eq.~(\ref{grBernoulli}) the 0PN- and 1PN-level Bernoulli equations.
The 0PN equation reads
\be
\label{0Bernoulli}
h_0+U_0-\frac{ \delta -1}{2(1+ \delta )} \Omega^2r^2  = c_0,
\ee
where $c_0$ is a constant that  can be interpreted as the energy per unit mass.
At the Newtonian level this is supplemented by the Poisson equation for the gravitational potential 
\be
\label{DeltaU0}
\Delta U_0 = 4\pi G   \rho_0,
\ee
where $\Delta$ denotes the flat Laplacian. The first correction $v^\phi_1$ to the angular velocity $\Omega $ is obtained from the perturbation expansion of the rotation law (\ref{rotation_law}) up to terms of the order $c^{-2}$. Assuming that  $\Psi_0 = 4 c_0$, one arrives at
 \begin{equation}
\label{constraint_solution_th}
v^\phi_1 = - \frac{2}{1 - \delta} \Omega_0^3 r^2 + \frac{A_\phi}{r^2 \left( 1- \delta \right)} - \frac{4\Omega_0 h_0}{1 - \delta },
\end{equation}
where we applied Eqs.~(\ref{zeroth_rotation_law}) and (\ref{0Bernoulli}).  

Remember that in the Newtonian gauge imposed in the line element (\ref{metric1}) the geometric distance to the rotation axis   is given by $\tilde r=r(1-U_0/c^2)+\mathcal{O}(c^{-4})$.   It is enlightening to write down the full expression for the angular velocity, up to the terms ${ \mathcal{O}(c^{-4})}$:
 \begin{eqnarray}
\label{angular velocityc2}
\Omega &=&\Omega_0+\frac{v^\phi_1}{c^2} =\frac{w}{\tilde r^{2/(1-\delta)}}    -\frac{2}{c^2(1 -  \delta )} \Omega_0 \left( U_0+\Omega_0^2r^2\right) \nonumber\\
&&\ + \frac{A_\phi}{r^2 c^2\left( 1- \delta \right)}   -  \frac{4}{c^2(1 -  \delta )} \Omega_0 h_0 .
\end{eqnarray}
This expression reduces to 
 \begin{eqnarray}
\label{angular velocityc2test}
\Omega &=&\Omega_0+\frac{v^\phi_1}{c^2} =\nonumber\\
&& \frac{w}{\tilde r^{2/(1-\delta)}}-  \frac{4}{rc^2(1 -  \delta )} \Omega_0 h_0  ,
\end{eqnarray}
in the case of test fluids, at the symmetry plane $z=0$. For the   dust, in the Schwarzschild geometry,  we get   
\begin{equation}
\label{angular velocityc2dust}
\Omega =\Omega_0+\frac{v^\phi_1}{c^2} =\frac{w}{\tilde r^{3/2}}   ;
\end{equation} 
the   1PN  correction    to $\Omega_0 $ is equal to $\frac{3U_0}{2c^2}\Omega_0$.  Thus the FC condition     is satisfied in the 1PN order.

After these consideration we are able to interpret the meaning of various contributions to the 1PN angular velocity $\Omega $. 
 The first term is simply the Newtonian rotation law rewritten as a function of the geometric distance, as given at the 1PN level of approximation, from the rotation axis. 
 The second term in (\ref{angular velocityc2}) vanishes at the plane of symmetry, $z=0$, for circular Keplerian motion of test fluids in the monopole  potential $-GM/R$.  Thus it is sensitive  both to the contribution of the disk self-gravity at the plane $z=0$ and the deviation  from the strictly Keplerian motion.
 The third  term is responsible for the geometric frame dragging. The last    term represents the recently discovered dynamic anti-dragging effect;  it  agrees (for the monomial angular velocities $\Omega_0 =r^{-2/(1- \delta )}w$) --- with the result obtained earlier in \cite{JMMP}.  

\textit{A comment on the term $-\frac{2}{c^2(1 -  \delta )}    \Omega_0^3r^2 $, that  has been missing in  \cite{JMMP}.} The reason for this omission is following. There  is a gauge freedom in choosing an integrability condition for the 1PN hydrodynamic equation; due to that the Bernoulli equation of the 1PN  order is specified up to a function $F(r)$. 
We assumed in \cite{JMMP}, in order to get the 1PN Bernoulli equation as in  \cite{BDS}, that $F(r)= 0$; but that is not consistent with  the   F-Condition. It appears that the right value is $F(r)=-\Omega_0^4r^4/(1+\delta)$, which leads to the emergence of the term in question.  

The vectorial component $A_\phi $ satisfies the following equation
\be
\label{Afi}
\Delta A_\phi -2\frac{\pa_rA_\phi }{r}= -16 \pi G r^2 \rho_0 \Omega_0 .
\ee   

 The 1PN Bernoulli equation does not influence the 1PN correction to the angular velocity. It has the form  
 \begin{eqnarray}
c_1 & = & -h_1 - U_1 -\Omega_0 A_\phi + 2 r^2 (\Omega_0)^2 h_0 - \frac{3}{2} h^2_0 \nonumber \\
& & - 4 h_0 U_0 - 2 U_0^2 - \frac{\delta -1}{4\left( 1+\delta \right)  } r^4\Omega^4_0 +F(r),
\label{Psi}
\end{eqnarray}
where $c_1$ is a constant. In order to derive (\ref{Psi}) we again used  $\Psi_0 =4c_0 $.  This result  agrees  with the 1PN calculation of  \cite{JMMP} up to the term $F(r)$.  
 
The 1PN potential correction  $U_1$ can be obtained from   
\begin{equation}
\label{DeltaU1}
\Delta U_1 = 4\pi G \left(   \rho_1 + 2p_0 + \rho_0(h_0-2U_0+2 r^2(\Omega_0)^2) \right).
\end{equation}

Equations (\ref{Afi}) and (\ref{DeltaU1}) have been derived in \cite{JMMP} in the framework of 1PN approximation. They can be also obtained directly from the Einstein equations written for the metric (\ref{metric}), as derived e.g.\ in \cite{komatsu}. Here we recall a version similar to that used in \cite{nishida_eriguchi}; it turns out to be more conveninent than the original form of \cite{komatsu}. The relevant equations read
\begin{eqnarray*}
\Delta \nu & = & 4 \pi \frac{G}{c^2} e^{2 \alpha/c^2} \left[ \rho (c^2 + h) \frac{1 + V^2/c^2}{1 - V^2/c^2} + 2 p \right]\\*
&& + \frac{1}{2 c^4} r^2 e^{2(\beta - \nu)/c^2} \nabla \omega \cdot \nabla \omega - \frac{1}{c^2} \nabla (\beta + \nu) \cdot \nabla \nu
\end{eqnarray*}
and
\begin{eqnarray*}
\left( \Delta + \frac{2}{r} \partial_r \right) \omega & = & - 16 \pi \frac{G}{c^2} e^{2 \alpha/c^2} \rho (c^2 + h) \frac{\Omega - \omega/c^2}{1 - V^2/c^2} \\*
& & + \frac{1}{c^2} \nabla (\nu - 3 \beta) \cdot \nabla \omega,
\end{eqnarray*}
where $\nabla$ denotes the ``flat'' gradient operator. The remaining Einstein equations yield corrections of higher orders.

{\bf In summary, we have shown that --- for $-\infty < w < \infty $ and $-\infty \le \delta \le 0$, $\delta \neq -1$ ---  the choice  $\kappa =(3-\delta )/(1+\delta ) + \mathcal{O}(c^{-2})$ and $\Psi =4c_0 + \mathcal{O}(c^{-2})$ in the rotation law (\ref{rotation_law})  guarantees that if there exists an exact solution analytic in powers of $c^{-2}$, then it satisfies the 0PN and   1PN approximating equations.  }

One  easily finds out that the rotation law (\ref{momentum}) satisfies the  generalized Rayleigh criterion  \cite{KEH89} for stability  $\frac{dj}{d\Omega }< 0 $  up to 1PN order, assuming that $\delta $ is strictly negative.

\textit{Comments on the 1PN corrections to the angular velocity.}  In the following considerations we assume $w>0$, which means $\Omega_0 >0$, but the reasoning is symmetric under the parity operation $w\rightarrow -w$.
The specific enthalpy  $h\ge 0$ is nonnegative, thence    $- \frac{4\Omega_0h_0}{1-\delta } $ is nonpositive --- the discovered in \cite{JMMP}   instantaneous 1PN dynamic  reaction slows  the motion: it ``anti-draggs'' a system.
 In contrast to that, the well known geometric term with $A_\phi$ is positive \cite{JMMP}, and the contribution $\frac{A_\phi}{r^2\left( 1 - \delta \right)}$ to the angular velocity is positive --- it pushes a rotating fluid body forward.  Thus  the two terms in  (\ref{constraint_solution_th}) counteract.  
 
Dust is special --- the  specific enthalpy $h_0$ vanishes, hence dust test bodies   are  exposed only to the geometric effect --- the frame dragging.  Even more special is the rigid (uniform) rotation --- the correction terms $v_1^\phi$ are proportional to $1/(1-\delta )$ and they vanish, because  now  $\delta = - \infty$. Uniformly rotating  disks  are already known to minimize the total mass-energy for a given baryon number and total angular momentum \cite{Boyer_Lindquist}. The vanishing of the 1PN correction $v_1^\phi $ is their another distinguishing feature.  

It follows from our discussion that  assuming the F-Condition, one has three free parameters: $w, \delta $ and $\Psi $; the parameter $\kappa $ is a given function of $\delta $. The full system of Einstein-Bernoulli equations can be solved numerically  within this class  of data and the resulting    solutions are expected to possess 0PN and 1PN limits.

\section{Concluding remarks}

We write down the general-relativistic rotation laws,  recover    the well known  geometric dragging of frames and derive a full form of  the two  other weak-fields  effects, including  the dynamic anti-dragging effect of  \cite{JMMP}.  The latter   can be robust according to the numerics of \cite{JMMP}, but the ultimate conclusion requires a fully general-relativistic treatment, that is the use of the new  rotation laws.    The   frame dragging occurs  --- through  the   Bardeen-Petterson effect \cite{Bardeen-Petterson} ---   in some AGN's \cite{Moran}. The two other effects can  lead    to its observable modifications    in  black hole systems with heavy  disks. 

In the weak field approximation of general relativity the angular velocity of toroids depends primarily on the distance from the rotation axis --- as in the Newtonian hydrodynamics --- but  the  weak fields contributions  make the rotation curve dependent on the height above the symmetry plane of a toroid.   
 
The new  rotation laws would allow   the investigation  of self-gravitating fluid bodies in the regime of strong gravity for general-relativistic versions of Newtonian rotation curves. 
In particular, they  can be used in order to describe stationary heavy  disks in tight accretion systems with central black holes. These highly relativistic systems  can be created in the merger of compact binaries consisting of pairs of black holes and neutron stars  \cite{Pan_Ton_Rez, Lovelace}, but they might   exist in some active galactic nuclei.  

The new general-relativistic rotation laws can be applied to the study of various open problems in   the post-Newtonian perturbation scheme of general-relativistic hydrodynamics. We demonstrate in this paper that an adaptation of the condition used in \cite{DJS} ensures uniqueness up the 1PN order. Further applications include the investigation of    convergence of the post-Newtonian pertubation scheme, as well  as the existence of  the Newtonian and post-Newtonian limits of   solutions.

\begin{acknowledgments}
PM  acknowledges the support of the  Polish Ministry of Science and Higher Education grant IP2012~000172 (Iuventus Plus). EM thanks Piotr Jaranowski for discussions on the PN approximations.
\end{acknowledgments}


\begin{thebibliography}{99}
\bibitem{Tassoul}
J.-L.\ Tassoul,
\textit{ Theory of rotating stars}
Princeton, N. J.:  Princeton University Press, 1978.
\bibitem{GYE}
F.\ Galeazzi, S.\ Yoshida, and Y.\ Eriguchi,
Astronomy and Astrophysics \ {\bf 541}, 156 (2012).
\bibitem{komatsu}
H.\ Komatsu, Y.\ Eriguchi, and I.\ Hachisu,
Mon.\ Not.\ R.\ Astron.\ Soc.\ {\bf 237}, 355 (1989).
\bibitem{Bardeen_1970}
J.\ M.\ Bardeen,
Astrophys.\ J.\ \textbf{162}, 71 (1970).
\bibitem{Butterworth_Ipser}
E. \ Butterworth and I.\ Ipser,
Astrophys.\ J.\ \textbf{200}, L103 (1969).
\bibitem{nishida_eriguchi}
S.\ Nishida and Y.\ Eriguchi,
Astrophys.\ J.\ {\bf 427}, 429 (1994).
\bibitem{nishida1}
S.\ Nishida, Y.\ Eriguchi, and A.\ Lanza,
Astrophys.\ J.\ {\bf 401}, 618 (1992).
\bibitem{DJS}   T.\ Damour, P.\ Jaranowski and G.\ Sch\"afer,
Phys.\ Rev.\  \textbf{62}, 044024 (2000).
\bibitem{MMP}  P.\ Mach, E.\ Malec and M.\ Pir\'og, Acta Phys.\ Pol.\ \textbf{ B44}, 107 (2013).
\bibitem{BDS}
L.\ Blanchet, T.\ Damour, and G.\ Sch\"afer,
Mon.\ Not.\ R.\ Astron.\ Soc.\ \textbf{242}, 289 (1990).
\bibitem{JMMP} P.\ Jaranowski, P.\ Mach, E.\ Malec, and M.\ Pir\'og,
    Phys.\ Rev.\ \textbf{D91}, 024039(2015).
\bibitem{KEH89} H.\ Komatsu, Y.\ Eriguchi, and I.\ Hachisu,
Mon.\ Not.\ R.\ Astron.\ Soc.\ {\bf 239}, 153 (1989).
\bibitem{Boyer_Lindquist}
R.\ H.\ Boyer and R.\ W.\ Lindquist,
Phys.\ Lett.\ \textbf{20}, 504 (1966).
\bibitem{Pan_Ton_Rez} F.\ Pannarale,  A.\ Tonita and   L.\ Rezzolla, 
Astrophys.\ J.\ \textbf{727},  art. id. 95(2011).
\bibitem{Lovelace} G.\ Lovelace, M. D.\ Duez, F.\ Foucart, L. E.\ Kidder, H. P.\ Pfeiffer, M. A.\ Scheel, B.\ Szilagyi, Class.\ Quantum\ Grav.\ \textbf{30},  art. id. 135004 (2013).
\bibitem{Bardeen-Petterson}
J.\ Bardeen and J.\ Petterson,
Astrophys.\ J.\ \textbf{195}, L65 (1975). 

\bibitem{Moran} 
J.\ Moran,   ASP Conference Series \textbf{395}, 87 (2008). 
 \end{thebibliography}
\end{document}